\documentclass[9pt,twocolumn,twoside]{osajnl}

\journal{ol} 

\setboolean{shortarticle}{true} 
\usepackage{graphicx}
\usepackage{dcolumn}
\usepackage{bm}
\usepackage{siunitx}
\usepackage{color}
\newcommand{\jj}{\mathrm{j}}
\newcommand\abs[1]{\left|#1\right|}
\newcommand{\E}{\mathrm{e}}

\title{Asymmetric surface wave excitation through metasurface-edge diffraction}

\author[1,*]{Miguel Camacho}
\author[2]{Filippo Capolino}
\author[3]{Matteo Albani}

\affil[1]{Department of Electrical and Systems Engineering, University of Pennsylvania, Philadelphia, Pennsylvania 19104, USA}
\affil[2]{Department of Electrical Engineering and Computer Science, University of California, Irvine, California 92697, USA}
\affil[3]{Department of Information Engineering and Mathematics, University of Siena, Siena, Italy}

\affil[*]{Corresponding author: mcamagu@seas.upenn.edu}

\ociscodes{240.6690, 260.1960, 310.2790, 310.6628}

 \doi{\url{https://www.doi.org/10.1364/OL.398342}}

\begin{abstract}
The selective excitation of localized surface wave modes remains a challenge in the design of both leaky-wave and bound-wave devices. In this Letter, we show how the truncation of a metasurface can play an important role in breaking the spatial inversion symmetry in the excitation of surface waves supported by the structure. This is done by combining a large anisotropy in the dispersion relation and the presence of an edge which also serves as a coupling mechanism between the plane wave excitation and the induced surface waves. By resorting to the exact solution of the scattering problem based on a discrete Wiener-Hopf technique we show that by inverting the component of the impinging wavevector parallel to the truncation, two distinct surface waves are excited.
\end{abstract}

\setboolean{displaycopyright}{true}

\begin{document}

\maketitle

Surface plasmons and their ability to propagate at an interface between a metal and free space have been a subject of intense research since their discovery by Ritchie \cite{Ritchie1957}. Mastering their excitation and propagation properties are ongoing challenges for the development of a new generation of optical devices ranging from optical circuitry to analog signal processing  \cite{Silva2014}.

Even before the attention was focused on surface plasmons, an analogous family of surface waves was largely investigated at much lower frequencies such as microwaves, at which metals do not behave as plasmas, but waves are supported by periodic metal-patterned structures \cite{ElliottSW,Cutler}. Also known as spoof-plasmons \cite{Pendry2004}, this type of surface waves allows for a minutious design of their propagation properties, defined in general by their complex dispersion relation. With the down-scaling of these patterned wave-supporting structures, metasurfaces, the flat counterpart of metamaterials, were devised \cite{Maci2011} with diverse applications \cite{Minatti2016, Dockrey2013}.


Surface waves cannot be excited directly by an incoming plane wave due to their wavevector mismatch, leading researchers to propose the use of evanescent fields associated with dielectric prisms under internal total reflection regime \cite{Otto1968} and also spatial harmonics generated by diffraction gratings \cite{Sambles1991}. Both these techniques require to place objects in the near field of the wave-supporting structure adding unwanted coupling mechanisms. Alternatively, localized sources placed in the vicinity of a metasurface are able to excite all wavector components due to their scattered-field's continuous spectrum \cite{Capolino2005, Camacho2017a}. However, the main drawback of this technique is the complete lack of selectivity in which surface waves are being excited.

\begin{figure}[htpb]
 \centering
 \includegraphics[trim=0cm 0cm 0cm 0cm, clip=true,width=0.79\columnwidth]{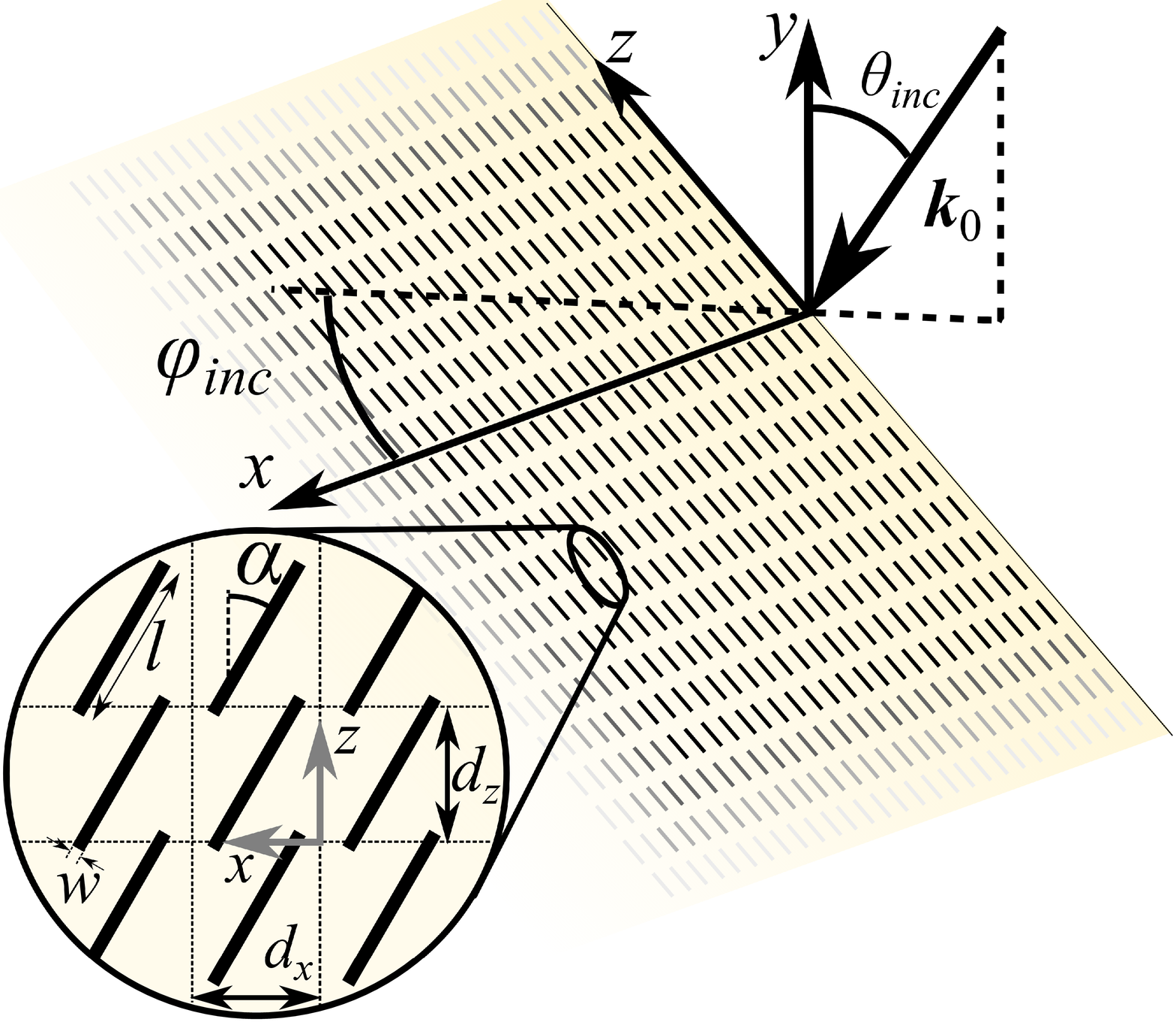}
 \caption{Truncated metasurface made of free standing narrow metallic patch dipoles rotated by an angle $\alpha$ with respect to the $z$ direction, under plane wave illumination.}
 \label{Fig_UC}
 \end{figure}
 
In this paper, we show that the special spectral characteristics of the scattering of a plane wave by the edge of a metasurface allows for a precise control of the excitation of surface waves. Thanks to the breakage of the spatial inversion symmetry due to the presence of the edge, one can excite two different surface waves when using plane waves with opposite signs of the wavevector component along the edge. Let us consider a simple metasurface consisting of a periodic array of narrow metallic flat patches in a rectangular lattice as those shown in Fig.\ref{Fig_UC}, which are rotated by an angle $\alpha$ with respect to the lattice vector along the $z$ axis. The semi-infinite, zero-thickness metasurface is truncated at $x=0$, and assumed to be infinitely long in the positive $x$ direction. This choice of the metasurface elements allows for a very accurate and simple mathematical solution for the fields, including the excited surface waves. This allows us to exploit the results in \cite{Camacho2019} to validate our rationale, although the choice of the geometry has no meaningful impact on the physics behind the selectivity of the surface wave excitation presented here.

As shown in \cite{Camacho2019}, under the approximation that each element of the metasurface is dominated by a realistic electric dipolar contribution, scattered fields and surface wave can be determined exactly. In other words, by assuming that the shape of the electric current on the surface of each dipole is known and described by a single basis function (that is a standard simplified approach in the Method of Moments \cite{Harrington93}), we obtain an exact solution for the value of the current on each of the elements of the semi-infinite array when this is illuminated by a plane wave with wavevector $(k_{x0}, k_{y0}, k_{z0})$. Since the array is infinite and periodic along the $z$ direction, all currents $i_{np}$ for $n=0,1,2,..$ and $p=0, \pm1, \pm2, ...$ can be simply represented as $i_{np}=i_{n}e^{-\jj k_{z0} p d_z}$, where $d_z$ is the array period along $z$. An $e^{j\omega t}$ time dependence is implicitly assumed and not shown. Hence, the current on the metasurface is fully represented by the succession $i_n$, obtained by solving the infinite series 
 
 \begin{equation}
\sum_{n=0}^\infty k_{m-n} i_n= Ve^{-\jj k_{x0} m d_x}, \quad(m=0,1,\dots,\infty)
\label{sys}
\end{equation}
where $k_{m-n}$ represents the electromagnetic coupling between two linear arrays of dipoles, both periodic along the $z$ direction, separated by a distance $d_x (m-n)$, and the right hand side is the source term obtained by the spatial overlap integral $V$ between the plane wave and the dipole current distribution, accounting for the phasing imposed by the impinging wave. Following \cite{Capolino2009,Camacho2019}, the exact solution of this infinite system is obtained analytically using the concept of the Z-transform into the variable $\xi$, leading to the total current on the $n$-th dipole expressed in the form of a closed-path integral in the complex $\xi$ spectral plane: 
\begin{equation}
i_n=\frac{V}{2\pi \jj}\frac{1}{K^-(\xi_\gamma)}\oint_C \frac{1}{K^+(\xi)}\frac{\xi^n}{\xi-\xi_\gamma} \text{d}\xi \label{ingen}
\end{equation}
The kernel functions $K^+(\xi)$ and $K^-(\xi_\gamma)$ are derived using the discrete Wiener-Hopf approach, as discussed in detail in \cite{Capolino2009,Camacho2019}, where $\xi$ is analogous to $k_x$ via the conformal mapping $\xi=e^{-\jj k_{x} d_x}$ and $\xi_\gamma=e^{-\jj k_{x0} d_x}$. In summary, first equation \eqref{sys} is represented in the Z-transformed domain, in terms of the transformed functions $K(\xi)$ and $I(\xi)$. Then, the spectral expression of the current $I(\xi)$ is found by factorizing $K(\xi)=K^+(\xi)K^-(\xi)$ in two spectral functions that have singularities in two distinct spectral regions. In particular $K^+(\xi)$ and $K^-(\xi)$ are free of zeroes and singularities outside and inside the complex $\xi$ unit circle, respectively. Finally, one reaches the expression in \eqref{ingen} by using the inverse Z-transform. 

The current integral in \eqref{ingen} is evaluated in terms of the spectral singularities inside the unit circle, that are a branch cut and poles whose contributions are calculated via their residues, as shown in \cite{Capolino2009,Camacho2019}. The procedure leads to the total current represented as the sum of three different wave species, i.e., three current terms originating from different physical mechanisms: (1) the residue of the pole at $\xi_\gamma$ is associated with the currents found in the non-truncated problem (double infinite periodic metasurface); (2) the branch cut introduced by the Green's function in the transform of the coupling coefficient $k_{m-n}$ is associated with the so-called "space wave" collecting the continuous wavevector spectrum of the fields diffracted by the edge of the metasurface; and (3) the residues of the poles arising from the zeroes inside the unit circle of the transform of the coupling $K(\xi)$ corresponding to the surface waves supported by the metasurface. 

 \begin{figure}[htpb] 
 \centering
 \includegraphics[trim=0cm 0cm 0cm 0cm, clip=true,width=0.75\columnwidth]{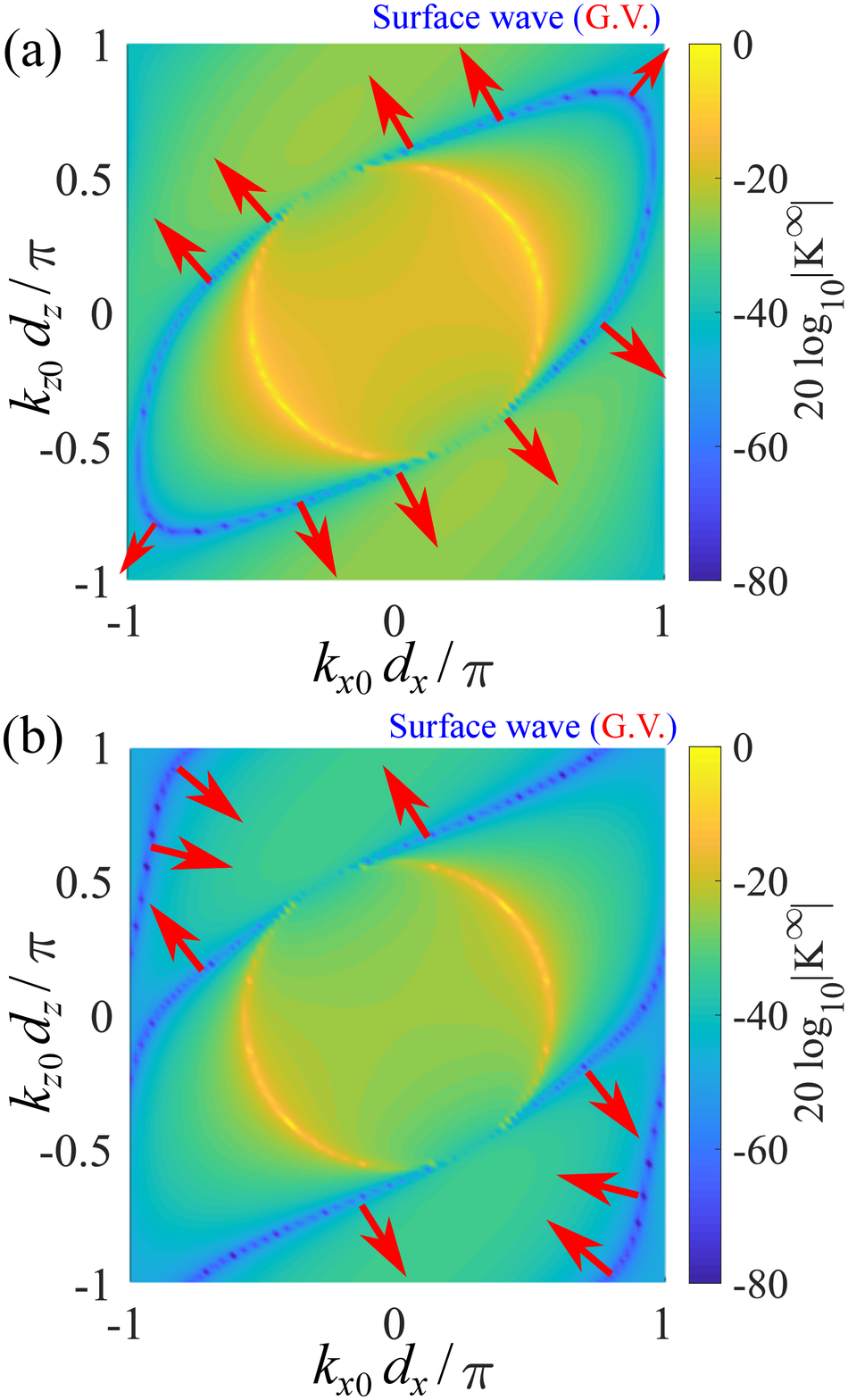} 
\caption{Isofrequency map of the spectrum $\abs{K^\infty}$ of the non-truncated array of dipoles rotated an angle of $\alpha=30^o$ from the $z$ axis, for the case of $d_x=d_z=\SI{3}{\milli\meter}$, $l=\SI{5}{\milli\meter}$ and $w=\SI{0.5}{\milli\meter}$ at (a) $\SI{27.3}{\giga\hertz}$ and (b) $\SI{28.6}{\giga\hertz}$. The $k_x, k_z$ pairs that provide vanishing values (i.e., $- \infty$ dB) represent the spectrum of surface waves and the red arrows the direction of their group velocity.}
 \label{Fig_kxy}
 \end{figure} 
 
The fact that not all of the surface waves supported by the structure are necessarily excited by the edge diffraction is an essential feature of the scattering by a semi-infinite structure, and this constitutes a significant difference with respect to the scattering by a small object/defect over or within the metasurface \cite{Camacho2017a}. Specifically, this is due to the coupling mechanism between the plane wave and the surface wave at the edge; only modes that propagate energy away from the edge are "physical", i.e., excitable by the diffracted wave originated by the edge or by a localized source \cite{Capolino2005,Campione2011}. This fact leads to the surprising phenomena such as the selective excitation of different sets of surface waves when the sign of the in-plane wavevector component of the plane wave is inverted. Remarkably, this phenomenon is independent of the phase velocities of the surface waves, as we demonstrate in the following.

\begin{figure}[htpb]
 \centering
 \includegraphics[trim=0cm 0cm 0cm 0cm, clip=true,width=0.84\columnwidth]{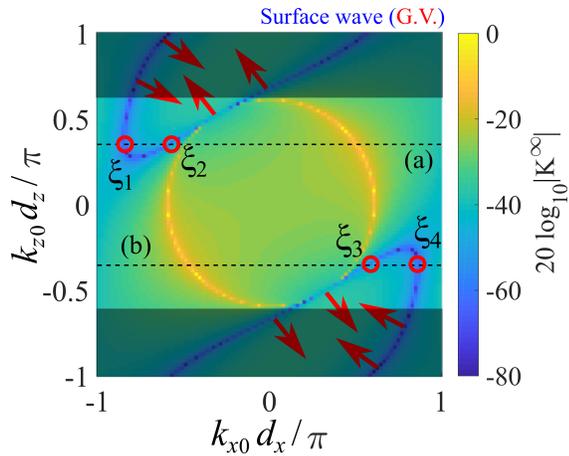} 
\caption{Isofrequency map of the spectrum $\abs{K^\infty}$ of the non-truncated array of dipoles rotated an angle of $\alpha=30^o$ for the case of $d_x=d_z=\SI{3}{\milli\meter}$, $l=\SI{5}{\milli\meter}$ and $w=\SI{0.5}{\milli\meter}$ at $\SI{30}{\giga\hertz}$. {Red arrows represent the direction of the group velocity of the surface waves}.}
 \label{Fig_kxy2}
 \end{figure}

The surface waves supported by a free standing array of narrow metallic dipoles tend to propagate in the direction perpendicular to the surface's dipole axis $\alpha$ as can be seen in Fig.~\ref{Fig_kxy} at two different frequencies. The surface wave isofrequency contour is the darkest blue region where $\abs{K^\infty}=0$, where $K^\infty\propto K(\xi_\gamma)$ represents the spectrum of the non-truncated metasurface assuming that both $k_{x0}$ and $k_{z0}$ vary. It is also found in Eq. (12) of \cite{Camacho2016}, here generalized to the case of oblique incidence. One can identify that Fig.~\ref{Fig_kxy}(a) is very similar to that of non-rotated dipoles presented in \cite{Camacho2019} with a rotation by the same angle as the dipoles. This is caused by the small coupling between the modes whose spectral information is visible in neighboring Brillouin zones. However as shown in Fig.~\ref{Fig_kxy}(b), as soon as the mode singularities touch the Brillouin zone boundary, they merge with their spatial harmonics, generating open contours. This change in behavior corresponds to the creation of bandgaps in the diagonal direction. Finally, as one further increases the frequency, the two pairs of open isofrequency contours on either side of the lightcone touch, forming the closed contours shown in Fig.~\ref{Fig_kxy2}.

A basic property of the isofrequency contours is that, given the time inversion symmetry imposed by Maxwell's equations, they must obey a point symmetry with respect to the origin, i.e., if $(k_x,k_z)$ represents a surface wave solution, so is $(-k_x,-k_z)$ with reversed phase and group velocities. Additionally, the periodicity of the reciprocal lattice imposes the fact that modes must cross the periodicity boundaries of the aforementioned zone in pairs, symmetrically positioned with respect to the center of the sides of the reciprocal lattice unit cell. Interestingly, this fact implies a high-symmetry degeneracy as these pairs collide in the corners when the dipoles are rotated by a certain angle. The corners, together with the center points of the sides of the Brillouin zone, are the only points where modes in neighboring Brillouin zones (i.e. spatial harmonics of the modes) are allowed to merge.

In contrast to the symmetric dispositions shown in \cite{Camacho2017a} where $(\pm k_x,\pm k_z)$ are all surface wave solutions, here we find that surface wave modes do not need to cross the Brillouin zone boundaries with zero group velocity (see the crossings in Fig.~\ref{Fig_kxy}(b) and the isofrequency contour further evolution in Fig.~\ref{Fig_kxy2}). Being able to achieve this type of behavior has been an intense topic of research in recent years, mostly associated with the use of higher symmetries \cite{Camacho2017c}, as it has important implications in the leaky-wave radiation angle. This non-zero group velocity can be explained by the asymmetry of the reciprocal lattice unit cell with respect to the $k_{x0}=0$ line, which means that the waves interfering at that boundary are not equal, i.e., there is lack of mirror symmetry with respect to the Brillouin zone boundary $k_{x0}=\pi/d_x$, therefore not constructing a standing wave, which is otherwise responsible for the zero group velocity found in metasurfaces whose unit cell is mirror-symmetric with respect to the $z-y$ plane. This lack of mirror symmetry in reciprocal space is therefore responsible for the excitation of two distinct waves when the impinging wavevector along the edge is reversed, as long as they are not excited at the exact crossing with the Brillouin zone boundary given by $k_{x0}=\pi/d_x$.

Having understood the evolution with frequency of the isofrequency contours associated with the surface modes supported by the array of tilted dipoles, let us now focus on how one can access these singularities {\it selectively} using edge diffraction. Due to the Floquet periodicity of the fields along the direction parallel to the edge, the wavevector spectrum of the problem is discrete: only surface modes with the same $k_z$ as the impinging plane can be excited. This limits the modes contained within $k_{z0}^\text{sw}<\abs{k_0}$ as limited by the dark regions in Fig.~\ref{Fig_kxy2}. Additionally, this means that once the wavevector of the plane wave is chosen, only the line dictated by $k_{z0}^\text{sw}=k_{z0}$ will be excited. Consequently, in Fig.~\ref{Fig_kxy2}, the excitation of surface waves will have a cut-off angle such that to be able to excite surface waves, $k_{z0}$ will need to be higher than the minimum of that of the isofrequency contour, which is equal to $k_{z0}^\text{sw}\approx\pm 0.25\pi/d_z$.

 \begin{figure}[htpb]
 \centering
 \includegraphics[trim=0cm 0cm 0cm 0cm, clip=true,width=0.75\columnwidth]{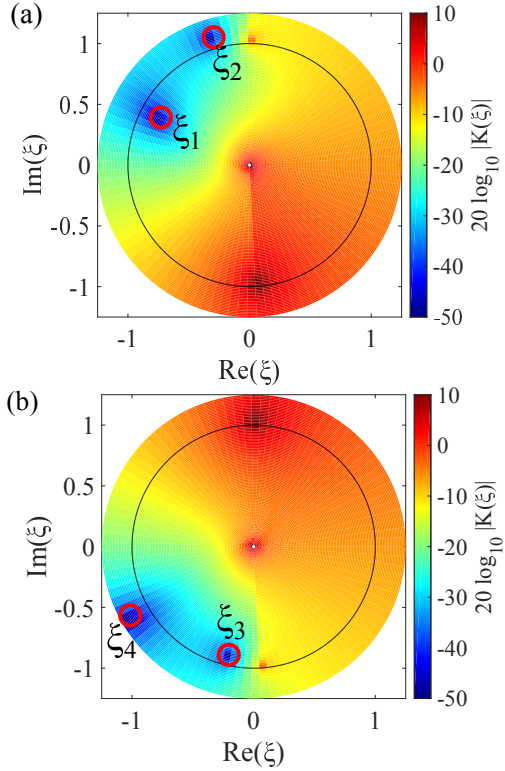} 
\caption{Map of the magnitude of $K(z)$ for Case (a) $k_{z0}=0.35 \pi/d_z$ and Case (b) $k_{z0}=-0.35 \pi/d_z$. The values of the other parameters are $d_x=d_z=0.3\lambda$, $\alpha=30^o$, $l=0.5\lambda$ and $w=0.05\lambda$. The solid line represents the unit circle. }
 \label{Fig_Kz_sw2}
 \end{figure}

Above the cut-off angle, the existence of surface waves can be seen in the complex $\xi$ plane as shown in Fig.~\ref{Fig_Kz_sw2} for the two cases: Case (a) $k_{z0}^\text{sw}=+ 0.35\pi/d_z$ and Case (b) $k_{z0}^\text{sw}=- 0.35\pi/d_z$. As pointed out earlier, the edge diffraction can only excite surface waves whose energy propagates according to a group velocity vector with positive $x$ component. Due to the isofrequency contour in Fig.~\ref{Fig_kxy2} being the result of the merging of contours from neighboring Brillouin zones, for Case (a) of $k_{z0}=0.35 \pi/d_z$ (depicted as dashed line (a)) one would expect the surface wave closer to the lightline ($\xi_2$) to originate from the first Brillouin zone, and therefore have a group velocity with negative $x$ component. Contrarily, the surface wave with the largest negative $k_{x0}^\text{sw}$ ($\xi_1$) arises from the next-to-the-left Brillouin zone and therefore propagates energy towards the right. Viceversa, when reversing the incident plane wave $z$ direction, i.e., for Case (b) when $k_{z0}=-0.35 \pi/d_z$, one concludes that the surface wave with smaller $k_{x0}^\text{sw}$ ($\xi_3$) originated in the first Brillouin zone, and therefore has positive group velocity along $x$ whilst the other ($\xi_4$) originated from the next-to-the-right Brillouin zone, and therefore would propagate its energy in the negative $x$ direction.

By accounting for the effect that losses have on the position of the surface wave spectral zeroes shown in Fig.~\ref{Fig_Kz_sw2}, we confirm the previous reasoning on the sign of the $x$ component of the group velocity of each mode. In Case (a) shown in Fig.~\ref{Fig_Kz_sw2}(a), due to the conformal transformation $\xi_{sw}=\E^{-\jj k_{x0}^{sw}d_x}$, a negative real part of $k_{x0}^{sw}$ means that moving along the solid line from $k_{x0}^{sw}=0$ to $k_{x0}^{sw}=\pi/d_x$ corresponds to the transformed coordinates following the unit circle anticlockwise from $\xi=1$ to $\xi=-1$. Therefore, the first zero, just outside the unit circle, corresponds to the wave originating in the first Brillouin zone with its energy propagating towards negative $x$ ($\xi_2$). However, the other surface wave appears just inside the unit circle, consistent with our reasoning for $\xi_1$. For Case (b), our deduction also holds as shown in Fig.~\ref{Fig_Kz_sw2}(b), in which case it is the surface wave with smaller $k_{x0}^{sw}$ ($\xi_3$) that transports energy towards positive $x$. As discussed earlier, only poles inside the unit circle correspond to "physical" (i.e. excitable) surface waves.

 \begin{figure}[htpb]
 \centering
 \includegraphics[trim=0cm 0cm 0cm 0cm, clip=true,width=0.83\columnwidth]{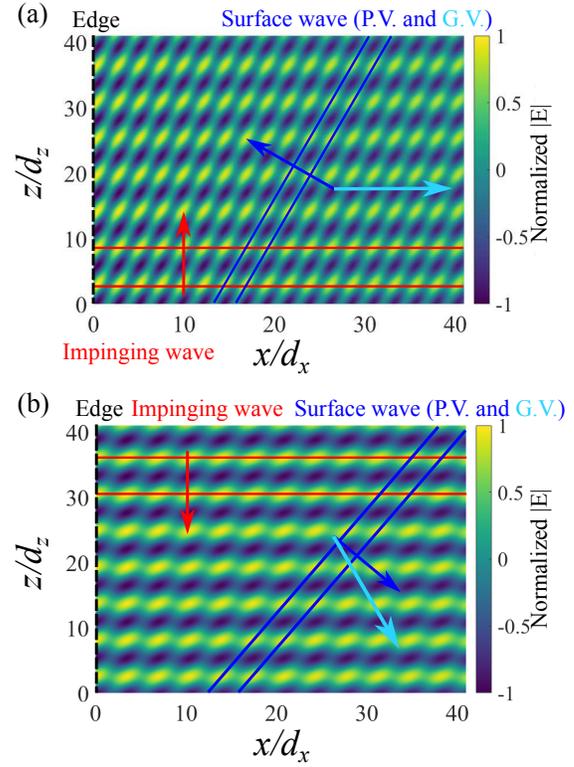} 
\caption{Normalized magnitude of the total current for the first 40x40 dipoles of the semi-infinite array from the truncation for the case of $d_x=d_z=0.3\lambda$, $\alpha=30^o$, $l=0.5\lambda$, $w=0.05\lambda$ $k_{x0}=0$ and (a) $k_{z0}=0.35 \pi/d_z$, and (b) $k_{z0}=-0.35 \pi/d_z$. }
 \label{Fig_Sw3}
 \end{figure}

Fig.~\ref{Fig_Sw3} presents the total current on the first 40 elements for the Cases (a) and (b) of Figs. \ref{Fig_kxy2} and \ref{Fig_Kz_sw2}. The red arrows represent phase velocity of the impinging plane wave while the red lines highlight its in-plane wavelength, and therefore also of the non-truncated-metasurface current contribution. The dark and light blue arrows represent the directions of the phase and group velocities of the surface wave, respectively. Notice that as discussed earlier, both surface waves present a positive group velocity component along $x$. It is also important to note the largely different surface-wave wavelengths in the two cases: (a) $\lambda_{sw}\approx0.67\lambda_0$ and (b) $\lambda_{sw}\approx\lambda_0$.

In conclusion, we have shown how the diffraction of a plane wave impinging on the edge of a metasurface constitutes a distinct coupling mechanism to selectively excite surface waves. Indeed, the combination of the semi-infinite nature of the problem and the presence of metasurface scatterers that are not symmetric with respect to the lattice vectors, allows the selective excitation of different sets of surface waves. The selected surface wave simply depends on the direction of the wavevector component of the impinging plane wave along the edge. This coupling selectivity (i.e., the "physical" excitation) is dictated by the group velocity of surface waves, which is understood from the merging of the isofrequency contours of the hybridizing modes.


\section*{Disclosures}
The authors declare no conflicts of interest.


\bibliography{library}








\bibliographyfullrefs{library}

\end{document}